\documentclass[preprintnumbers,10pt,nofootinbib]{revtex4}

\usepackage{amsmath,latexsym,amssymb,amsfonts}
\usepackage{graphicx,color}
\usepackage{bm}
\usepackage{microtype}
\usepackage{amsmath}
\usepackage{amssymb}
\usepackage{subfigure}
\usepackage{epstopdf}

\begin{document}

%\preprint{0000-00-00}

\title{Violation of weak cosmic censorship by the Oppenheimer-Snyder collapse}

\author{Il Gyeong Choi$^{a}$\footnote{{\tt eogkrrksek12@gmail.com}} and Dong-han Yeom$^{a,b,c,d}$\footnote{{\tt innocent.yeom@gmail.com}}
}

\affiliation{
$^{a}$Department of Physics Education, Pusan National University, Busan 46241, Republic of Korea\\
$^{b}$Research Center for Dielectric and Advanced Matter Physics, Pusan National University, Busan 46241, Republic of Korea\\
$^{c}$Leung Center for Cosmology and Particle Astrophysics, National Taiwan University, Taipei 10617, Taiwan\\
$^{d}$Department of Physics and Astronomy, University of Waterloo, Waterloo, ON N2L 3G1, Canada 
}

\begin{abstract}
We consider the possibility that the weak cosmic censorship conjecture can be violated using the Oppenheimer-Snyder collapse model with a perfect fluid star interior. Metric models with a naked singularity can be used, and the Oppenheimer-Snyder collapse might be possible; the null energy condition is also satisfied in these models. However, this is just a toy model because no known model as a solution to the Einstein equation. To avoid this problem, we can consider a naked singularity solution as a proper solution of the Einstein equation, but in this case, we need to introduce a thin-shell on top of the perfect fluid star. In this case, gravitational collapse is allowed, but the null energy condition should be violated at the thin-shell. In conclusion, we could not find a physically viable model that violates weak cosmic censorship, but it opens a window to study the properties of cosmic censorship constructively. The violation of weak cosmic censorship might be possible if a modified gravity model provides a solution that allows the Oppenheimer-Snyder collapse without a shell. Also, regarding the thin-shell case, the shell approaches the naked singularity closer without violating the null energy condition, although it must violate the null energy condition at the singularity eventually; this might be regarded as an effective violation of cosmic censorship in some sense.
\end{abstract}

\maketitle

\newpage

\tableofcontents

\section{Introduction}

The singularity is where we can see the limitations of general relativity \cite{Hawking:1970zqf}. Can we form a naked singularity using a gravitational collapse? If we can make it, what will its effect be? In almost all cases, forming such a naked singularity is impossible. This was conjectured by Penrose, so-called the \textit{cosmic censorship conjecture} \cite{Penrose:1969pc}. In this paper, we will focus on the weak version of the conjecture: every singularity derived from gravitational collapse must be hidden by an event horizon. It is fair to say that we do not have a generic proof of this conjecture, but we can learn many secrets in nature as we \textit{fail} to form a naked singularity.

Interestingly, finding a naked singularity solution as a mathematical solution of the Einstein equation is not difficult. The simplest example is a charged black hole with a greater charge than its mass \cite{RN}. The problem is that to make such a naked solution from gravitational collapse, there appears to be repulsions of charges, and therefore, the repulsive force eventually pushes charges outside; finally, an event horizon is formed and prevents the violation of the cosmic censorship conjecture; see examples for cases of four dimensions~\cite{Hong:2008mw}, three dimensions~\cite{Hwang:2011mn}, or different topologies~\cite{Hansen:2013vha}.

What happens if we consider a naked solution that has no electric charge? To deal with this issue, we meet the following two questions:
\begin{itemize}
\item[--] How can we reasonably design the gravitational collapse model?
\item[--] What background solution has a naked singularity? Is it an exact solution of the Einstein equation and matter field equations?
\end{itemize}
The first question means that if we do not properly design the gravitational collapsing model and rely on an approximate model only, we may see the formation of the naked singularity, but this is a misleading result.

To avoid this ambiguity, this paper uses two approaches. First, we use the \textit{Oppenheimer-Snyder} (OS) \textit{collapse model} \cite{Oppenheimer:1939ue}, which means we think the outside is a static solution, while a perfect fluid fills the inside. Hence, the star interior can be described as a Friedmann-Robertson-Walker (FRW) spacetime. However, as one thinks of the OS model, we meet the second question about a naked singularity solution; the application of the OS model has some limitations, since this model cannot cover the generic classes of static and spherical symmetric solutions. Therefore, if we consider gravitational collapse of a generic class of spherically symmetric solutions, we must extend the OS model with a thin-shell at the junction \cite{Israel:1966rt}. Hence, our second approach to describe gravitational collapse is introducing a thin-shell.

In this paper, we consider the OS collapse model and discuss whether this collapsing model can form a naked singularity. First, we consider the situation where a metric has a naked singularity, and the OS collapse is possible without a shell. In this case, a naked singularity \textit{can} be formed without violating any physical laws (e.g., causality or energy conditions \cite{Chew:2024ggd}), but the only problem is that there is no known background solution as a solution of the Einstein equation. To avoid this problem, we consider a background metric as a solution of the Einstein and matter field equations; we introduce the Janis-Newman-Winicour (JNW) solution as an example \cite{Janis:1968zz}. In this case, to form a naked singularity, the star's interior does not necessarily violate the energy condition, but the \textit{shell must violate the null energy condition} at some moment.

This paper is organized as follows. In Sec.~\ref{sec:gen}, we first review the OS collapse model, where we first consider the original version and second the case with a thin-shell. In Sec.~\ref{sec:nak}, we discuss naked singularity models. We first mention physical conditions of naked singularity models; later, we provide a metric ansatz as well as the JNW solution as an exact solution. In Sec.~\ref{sec:nf}, we evaluate the naked singularity formation models based on the OS collapse with or without thin-shells. Finally, in Sec.~\ref{sec:dis}, we summarize our results and meanings, and discuss possible future research topics.

\section{\label{sec:gen}General formalism of the Oppenheimer-Snyder collapse}

\subsection{Original version without a shell}

The OS collapse model describes a massive star's gravitational collapse, leading to a black hole's formation under an idealized condition \cite{Oppenheimer:1939ue}. This model assumes a spherical distribution of matter with a perfect fluid \cite{Poisson:2009pwt}. Motivated by this scenario, let us consider outside the collapsing star geometry $ds^{2}_{+}$ and inside the collapsing star geometry $ds^{2}_{-}$ are
\begin{eqnarray}
ds^{2}_{+} &=& - f(r) dt^{2} + \frac{1}{f(r)} dr^{2} + r^{2}d\Omega^{2},\\
ds^{2}_{-} &=& - dT^{2} + a^{2}(T) \left( d\chi^{2} + \sin^{2} \chi d\Omega^{2} \right),
\end{eqnarray}
where $f(r)$ is a function of $r$ and $a(T)$ is the scale factor of the interior region; the coordinate variables for outside are $[t, r, \theta, \varphi]$ and that of inside are $[T, \chi, \theta, \varphi]$. The junction between outside and inside $ds^{2}_{\Sigma}$ is
\begin{eqnarray}
ds^{2}_{\Sigma} &=& -d\tau^{2} + a^{2}(\tau) \sin^{2} \chi_{0} d\Omega^{2} \\
&=& \left( - f(r) \dot{t}^{2} + \frac{\dot{r}^{2}}{f(r)} \right) d\tau^{2} + r^{2}(\tau) d\Omega^{2},
\end{eqnarray}
where $\tau$ is the proper time of the junction, $\dot{}$ is the derivative for $\tau$, and $\chi_{0}$ is an arbitrary angle value, while its meaning will be specified later.

First, to impose the continuity between metrics, we require
\begin{eqnarray}
r(\tau) &=& a(\tau) \sin \chi_{0},\\
f(r) \dot{t}^{2} - \frac{\dot{r}^{2}}{f(r)} &=& 1.
\end{eqnarray}
Second, to impose the continuity of extrinsic curvatures $K^{\mu}_{\;\nu}$ between outside and inside, we require
\begin{eqnarray}
\left. K^{\tau}_{\;\;\tau} \right|_{-} &=& \left. K^{\tau}_{\;\;\tau} \right|_{+},\\ 
\left. K^{\theta}_{\;\;\theta} \right|_{-} &=& \left. K^{\theta}_{\;\;\theta} \right|_{+},
\end{eqnarray}
where
\begin{eqnarray}
\left. K^{\tau}_{\;\;\tau} \right|_{-} &=& 0,\\
\left. K^{\tau}_{\;\;\tau} \right|_{+} &=& \frac{\ddot{r} + \frac{f'}{2}}{\sqrt{\dot{r}^{2} + f}},\\
\left. K^{\theta}_{\;\;\theta} \right|_{-} &=& \frac{1}{a} \cot \chi_{0},\\
\left. K^{\theta}_{\;\;\theta} \right|_{+} &=& \frac{1}{r} \sqrt{\dot{r}^{2} + f}.
\end{eqnarray}

By plugging equations, we obtain the following relations:
\begin{eqnarray}
\sqrt{\dot{r}^{2} + f} &=& \cos \chi_{0}, \label{eq:1}\\
a &=& \frac{r}{\sin \chi_{0}}, \label{eq:2}\\
\dot{t}^{2} &=& \frac{1}{f} \left( 1 + \frac{\dot{r}^{2}}{f}\right).  \label{eq:3}
\end{eqnarray}
Hence, from Eq.~(\ref{eq:1}), one can solve $r(\tau)$; from this, we obtain $a(\tau)$ using Eq.~(\ref{eq:2}), and finally, we can determine $t(\tau)$ using Eq.~(\ref{eq:3}). These equations determine the system entirely.

One can further compute information about the interior region:
\begin{eqnarray}
\rho &=& \frac{3}{8\pi} \left( \frac{\dot{a}^{2} +1}{a^{2}} \right)\\
&=& \frac{3}{8\pi} \left( \frac{\dot{r}^{2} + \sin^{2}\chi_{0}}{r^{2}} \right),\\
p &=& - \frac{a \dot{\rho}}{3 \dot{a}} - \rho \\
&=& - \frac{r}{3} \frac{d\rho}{dr} - \rho,
\end{eqnarray}
where $\rho$ and $p$ are the energy density and the pressure of the perfect fluid region, respectively.

\subsection{Necessity of a thin-shell for generalized spherical symmetric and static metrics}

Now, let us generalize the OS collapse to more generic situations. If we consider a more general static and spherically symmetric spacetime, it is as follows:
\begin{eqnarray}
ds_+^{2} &=& - e^{-2\Phi(r)}  f(r) dt^{2} + \frac{1}{f(r)} dr^{2} + r^{2} d\Omega^{2},\\
ds_-^{2}&=& - dT^{2} + a^2(T) \left( d\chi^2 + \sin^2\chi d\Omega^{2} \right),\\
ds_\Sigma^{2} &=& -d\tau^2+a^2(t) \sin^2\chi_0 d\Omega^{2}\\
&=& \left( -e^{-2\Phi(r)}  f(r) \dot{t}^{2} + \frac{\dot{r}^{2}}{f(r)} \right) d\tau^{2} + r^{2} d\Omega^{2},
\end{eqnarray}
where $\Phi(r)$ is a function of $r$.

The following relationship is derived from the continuity of metric components at the boundary of two different metrics and the continuity of extrinsic curvatures:
\begin{eqnarray}
r(\tau)&=&a(\tau) \sin\chi_0,\\
 e^{-2\Phi}  f(r) \dot{t}^{2} - \frac{\dot{r}^{2}}{f(r)}  &=&1,\\
 \frac{\ddot{r} + \frac{f'}{2}}{\sqrt{\dot{r}^{2} + f}}+\Phi'{\sqrt{\dot{r}^{2} + f}}&=&0,\\
 \frac{1}{r} \sqrt{\dot{r}^{2} + f} - \frac{1}{a} \cot \chi_{0}&=& 0,
 \end{eqnarray}
where $'$ is a derivative for $r$.

After simple computations, it is easy to show that $\Phi'=0$ must be satisfied; otherwise, assigning the time evolution along the shell is impossible. Therefore, it is inevitable to conclude that the original OS collapse cannot be applied to the generic spherically symmetric and static spacetime; we must introduce a shell at the junction surface.

\subsection{Junction equations for a thin-shell}

The energy-momentum tensor of the shell we want to consider is expressed as follows \cite{Brahma:2018cgr}:
\begin{eqnarray}
-8\pi {S}^{a}_{\;b} &=&\left[{K}^{a}_{\;b}\right]-\left[K\right]{h}^{a}_{\;b},
\end{eqnarray}
where ${S}^{a}_{\;b}$ is the energy-momentum tensor and ${h}^{a}_{\;b}$ is an induced metric of the shell. The components of the energy-momentum tensor are as follows:
\begin{eqnarray}
4\pi S^{\tau}_{\;\tau}&=& \left.K^{\theta}_{\;\theta}\right|_+-\left.K^{\theta}_{\;\theta}\right|_- \\
&=& \frac{1}{r} \sqrt{\dot{r}^{2} + f}-\frac{1}{a}\cot \chi_{0},\\
8\pi S^{\theta}_{\;\theta}&=& \left.K^{\tau}_{\;\tau}\right|_+ -\left.K^{\tau}_{\;\tau}\right|_- + \left.K^{\theta}_{\;\theta}\right|_+ -\left.K^{\theta}_{\;\theta}\right|_- \\
&=& \frac{\ddot{r} + \frac{f'}{2}}{\sqrt{\dot{r}^{2} + f}}+\Phi'{\sqrt{\dot{r}^{2} + f}}+\frac{1}{r} \sqrt{\dot{r}^{2} + f}-\frac{1}{a}\cot \chi_{0},
\end{eqnarray}
where we can identify $-S^{\tau}_{\;\tau}$ as the tension of the shell ($\sigma$) and $S^{\theta}_{\;\theta}$ as the pressure of the shell ($\lambda$). From these equations, one can derive the following relations:
\begin{eqnarray}
\cos \chi_{0} - \sqrt{\dot{r}^{2} + f} &=& 4\pi r \sigma, \label{eq:shell1}\\
\sigma' &=& - \frac{2}{r} \left( \lambda + \sigma \left( 1 + \frac{\Phi'}{2} r \right) - \frac{\Phi'}{8 \pi} \cos \chi_{0} \right). \label{eq:shell2}
\end{eqnarray}
It is easy to check that these equations are consistent with more generic thin-shell junction equations (see cases for time-like shells \cite{Garcia:2011aa} and space-like shells \cite{Bouhmadi-Lopez:2020oia}).

\section{\label{sec:nak}Naked singularity models}

\subsection{Necessary conditions}

We will consider a metric with a naked (time-like) singularity in an asymptotically flat spacetime. The general spherical symmetric and static metric form is as follows:
\begin{eqnarray}
ds^{2} = - e^{-2 \Phi(r)}f(r) dt^{2} + \frac{1}{f(r)} dr^{2} + r^{2} d\Omega^{2},
\end{eqnarray}
where
\begin{eqnarray}
f(r) = 1 - \frac{2M(r)}{r},
\end{eqnarray}
and $M(r)$ is the Misner-Sharp mass \cite{Misner:1964je}. The physically necessary conditions are as follows:
\begin{itemize}
\item[--] \textit{Asymptotic flatness}: Physically we require that $f(r) \rightarrow 1$ and $\Phi(r) \rightarrow \mathrm{const.}$ in the $r \rightarrow \infty$ limit.
\item[--] \textit{No horizon}: $f(r) > 0$ for all $r$ is required to avoid the apparent horizon.
\item[--] \textit{Naked singularity at $r = 0$}: If $M(r) \sim r^{n}$, near $r \sim 0$, the Kretschmann scalar near $r = 0$ is approximately
\begin{eqnarray}
R_{abcd}R^{abcd} \propto \frac{r^{2n}}{r^{6}},
\end{eqnarray}
and hence, to have a singularity, we require $n < 3$ \cite{Nielsen:2008kd}.
\item[--] \textit{Null energy condition}: We require the null energy condition.
\end{itemize}

\subsection{Just an ansatz: a naked singularity model}

First, we consider the following form of the metric:
\begin{eqnarray}\label{eq:metricform}
ds^{2} = - f(r) dt^{2} + \frac{1}{f(r)} dr^{2} + r^{2} d\Omega^{2},
\end{eqnarray}
where
\begin{eqnarray}
f(r) = 1 - \frac{2M(r)}{r},
\end{eqnarray}
and $M(r)$ is the Misner-Sharp mass. This kind of ansatz was widely used, especially in the context of regular black hole models; for example, see \cite{Dymnikova:1992ux}.

We provide a model with the following physical properties:
\begin{eqnarray}
M(r) = \frac{2 M_{0}}{\pi} \tan^{-1} \frac{r}{r_{0}},
\end{eqnarray}
where $M_{0}$ and $r_{0}$ are constants. For a small $r$ region, $M(r) \sim r$, hence, a naked singularity is guaranteed. By choosing proper constants, it is not difficult to make $f(r) > 0$ for all $r > 0$. Also, it is easy to check that the asymptotic mass is $M_{0}$, and the asymptotic flatness condition is well satisfied. The null energy condition is satisfied if
\begin{eqnarray}
\frac{2M'}{r} - M'' > 0.
\end{eqnarray}
One can compute that
\begin{eqnarray}
\frac{2M'}{r} - M'' = \frac{4 M_{0}}{\pi (r/r_{0})} \frac{1 + 2 (r/r_{0})^{2}}{\left(1 + (r/r_{0})^{2}\right)^{2}}.
\end{eqnarray}
Hence, we can check that this metric ansatz satisfies all our required properties.

\subsection{An exact solution example: Janis-Newman-Winicour solution}

The JNW metric solves the Einstein-scalar field equations \cite{Janis:1968zz}, representing a spacetime with a massless scalar field $\varphi$. This metric inherently contains a naked singularity, which, unlike a black hole, lacks an event horizon, making the singularity directly observable from the outside. The metric of JNW is expressed as follows:
\begin{eqnarray} \label{JNW1}
ds^{2} = - f(R) dt^{2} + \frac{1}{f(R)} dR^{2} + r^{2} d\Omega^{2},
\end{eqnarray}
where
\begin{eqnarray}
f(R) &=& \left( \frac{2R - r_{0} (\mu-1)}{2R + r_{0} (\mu+1)} \right)^{1/\mu}, \\
r^{2} &=& \frac{1}{4} \left( 2R + r_{0} (\mu+1) \right)^{1+1/\mu} \left( 2R - r_{0} (\mu-1) \right)^{1-1/\mu}, \\
\varphi(R) &=& \frac{A}{\mu} \ln \left| \frac{2R - r_{0} (\mu-1)}{2R + r_{0} (\mu+1)} \right|,\\
\mu &\equiv& \left( 1 + \frac{4\kappa A^{2}}{r_{0}^{2}} \right) \geq 1,\\
\kappa &\equiv& 8\pi,\\
r_{0} &\equiv& 2m,
\end{eqnarray}
where $m$ is related to the ADM mass and $A$ is the charge of the scalar field (see more detailed information in~\cite{Chew:2024ggd}).

\section{\label{sec:nf}Naked singularity formation via the Oppenheimer-Snyder collapse}

\subsection{Oppenheimer-Snyder collapse without a shell}

\begin{figure}
\centering
\mbox{
(a)
\includegraphics[scale=0.6]{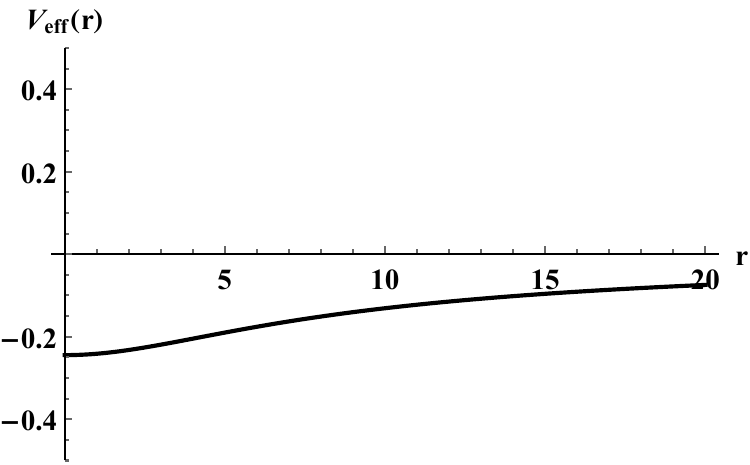}
(b)
\includegraphics[scale=0.6]{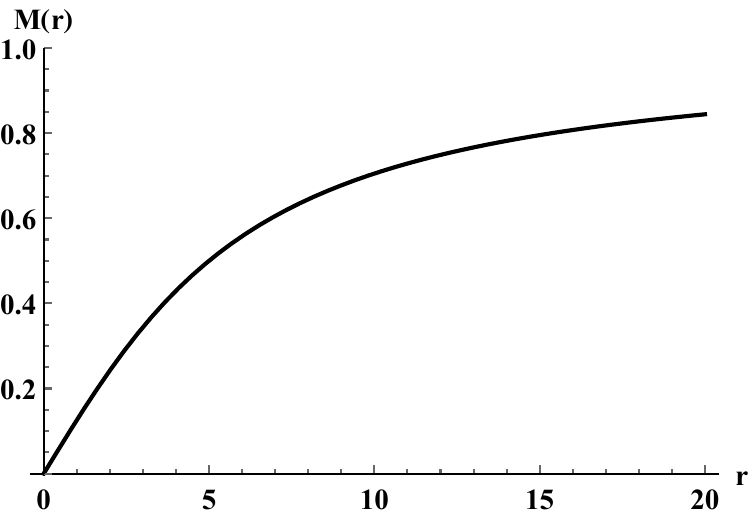}
}
\mbox{
(c)
\includegraphics[scale=0.6]{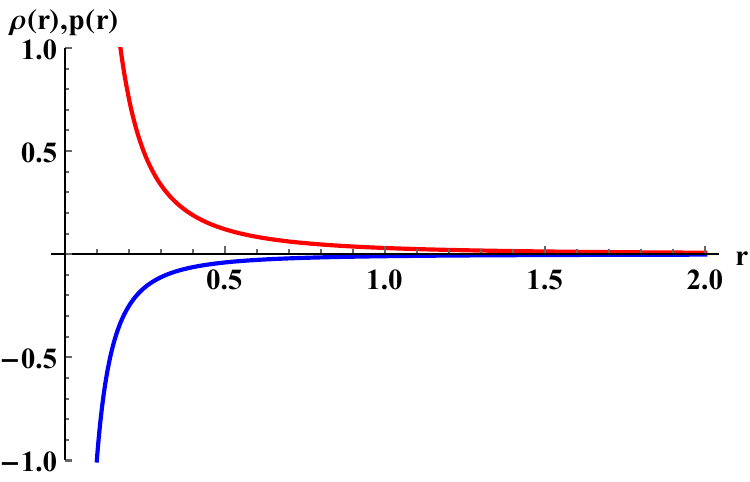}
(d)
\includegraphics[scale=0.6]{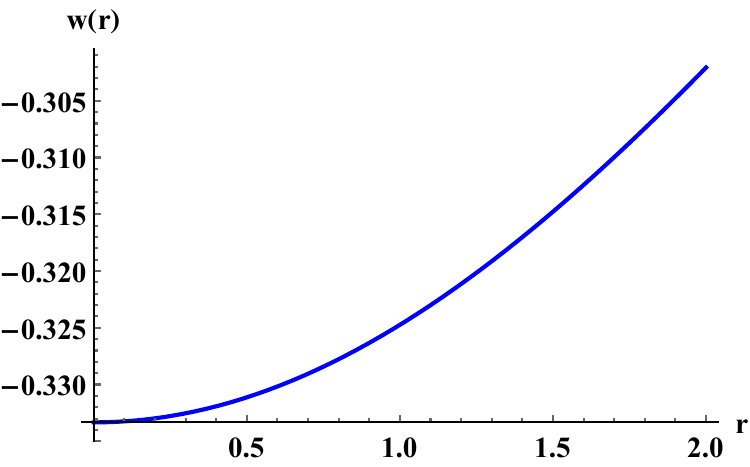}
}
\caption{\label{fig:type1}An example of the OS collapse without a shell, where $M_{0} = 1$, $r_{0} = 5$, and $\chi_{0} = 0.1$: (a) $V_{\mathrm{eff}}(r)$, (b) $M(r)$, (c) $\rho(r)$ (red) and $p(r)$ (blue), (d) $w(r)$.}
\end{figure}

Let us first consider the OS collapse without a shell using the metric form of Eq.~(\ref{eq:metricform}). The dynamics of $r$, Eq.~(\ref{eq:1}), is rewritten by
\begin{eqnarray}
\dot{r}^{2} + V_{\mathrm{eff}}(r) = 0,
\end{eqnarray}
where
\begin{eqnarray}
V_{\mathrm{eff}}(r) = f(r) - \cos^{2} \chi_{0}.
\end{eqnarray}
In Fig.~\ref{fig:type1}, its numerical demonstration is presented.

The classical trajectory is allowed only if $V_{\mathrm{eff}}(r) < 0$. Hence, to allow a gravitational collapse to the center, we require the condition that $V_{\mathrm{eff}}(r) < 0$, or equivalently,
\begin{eqnarray}
\sin^{2} \chi_{0} < \frac{2M(r)}{r}
\end{eqnarray}
for all $0 \leq r \leq r_{\mathrm{max}}$, where $r_{\mathrm{max}}$ can be an arbitrary positive value (including infinity).

Interestingly, if we consider our toy model
\begin{eqnarray}
M(r) = \frac{2 M_{0}}{\pi} \tan^{-1} \frac{r}{r_{0}},
\end{eqnarray}
around $r \sim 0$, one can expand
\begin{eqnarray}
\frac{2M(r)}{r} \simeq \frac{4 M_{0}}{\pi r_{0}},
\end{eqnarray}
and hence, by adjusting parameters, it is, in principle, possible to satisfy the condition $V_{\mathrm{eff}}(r) < 0$ at $r \sim 0$ (see Fig.~\ref{fig:type1} (a) and (b)). Therefore, the star interior can collapse to the center and form a naked singularity.

As we observed, our metric ansatz satisfies the null energy condition. Then, how about the energy-momentum tensor of the interior? Using a simple computation, the energy density $\rho$ and the pressure $p$ of the interior region are
\begin{eqnarray}
\rho &=& \frac{3}{8\pi} \left( \frac{2M(r)}{r^{3}} \right),\\
p &=& - \frac{M'}{4\pi r^{2}}.
\end{eqnarray}
Therefore, if $M(r) > 0$, then $\rho > 0$ for all $r$. One more necessary condition for the null energy condition of the star's interior is \cite{Poisson:2009pwt}
\begin{eqnarray}
w \equiv \frac{p}{\rho} \geq -1,
\end{eqnarray}
where
\begin{eqnarray}
w = - \frac{rM'}{3M}.
\end{eqnarray}
In our specific model, we obtain
\begin{eqnarray}
w = \frac{-r/(3r_{0})}{\tan^{-1} (r/r_{0}) \left( 1 + (r/r_{0})^{2} \right)},
\end{eqnarray}
which is greater than $-1$. As an example, see Fig.~\ref{fig:type1} (c) and (d).

Indeed, this is not an accident. Let us think about the marginal condition for the null energy condition:
\begin{eqnarray}
w = - \frac{rM'}{3M} = -1.
\end{eqnarray}
Hence, $3M = rM'$. If we differentiate both sides, we obtain $rM'' = 2M'$, but this is precisely the marginal condition of the null energy condition of the original metric. Hence, we can conclude that if the original metric satisfies the null energy condition, the corresponding star interior also satisfies the null energy condition.

\subsection{Oppenheimer-Snyder collapse with a shell}

Next, let us generalize the computations. We assume that the outside metric is the JNW solution. The dynamics of $r$, Eq.~(\ref{eq:shell1}), is rewritten by
\begin{eqnarray}
\dot{r}^{2} + V_{\mathrm{eff}}(r) = 0,
\end{eqnarray}
where
\begin{eqnarray}
V_{\mathrm{eff}}(r) = f(r) - \left( \cos \chi_{0} - 4\pi r \sigma \right)^{2}.
\end{eqnarray}
In addition to this, we need to solve the equation for the tension, Eq.~(\ref{eq:shell2}), where it is reasonable to assume the equation of state of the shell $w_{s} \equiv \lambda/ \sigma$ and solve the equation:
\begin{eqnarray}
\sigma' = - \frac{2}{r} \left( \left( 1 + w_{s} + \frac{\Phi'}{2} r \right) \sigma - \frac{\Phi'}{8 \pi} \cos \chi_{0} \right).
\end{eqnarray}
The null energy condition of the shell implies that $\sigma \geq 0$ and $w_{s} \geq -1$.

\begin{figure}
\centering
\mbox{
(a)
\includegraphics[scale=0.4]{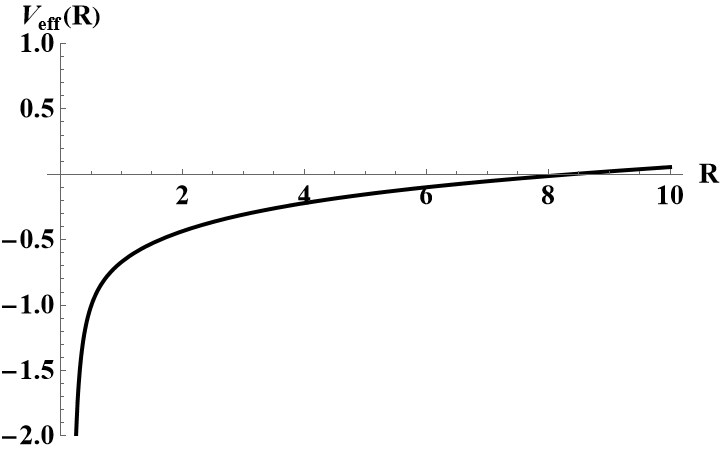}
(b)
\includegraphics[scale=0.4]{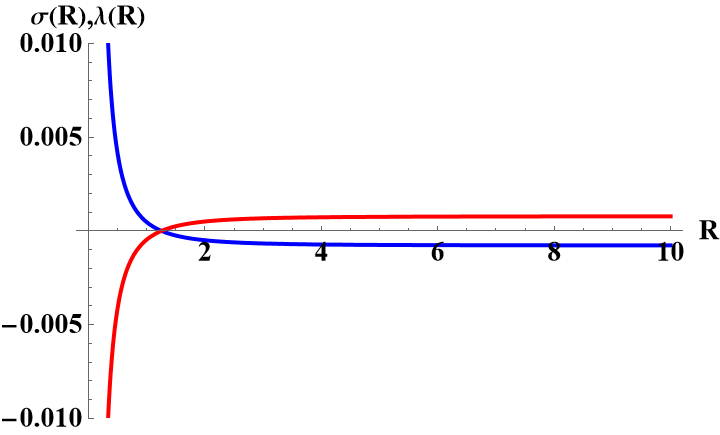}
}
\mbox{
(c)
\includegraphics[scale=0.4]{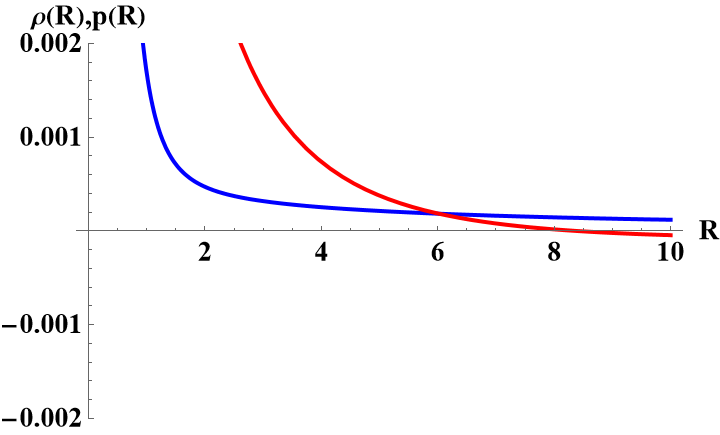}
}
\mbox{
(d)
\includegraphics[scale=0.4]{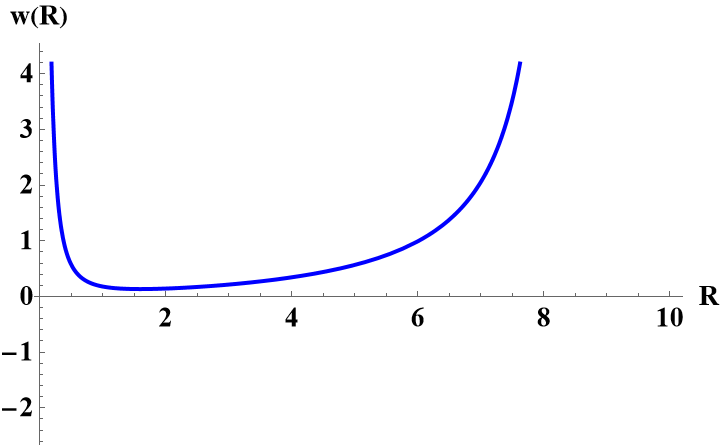}
}
\caption{\label{fig:type2}An example of the OS collapse with a shell assuming the JNW solution, where $A = 0.1$, $m = 1$,  $\sigma_{0} = 0.015$, $\chi_{0} = 0.01$ and $w_{s}=-1$: (a) $V_{\mathrm{eff}}$, (b) $\sigma$ (red), $\lambda$ (blue), (c) $\rho$ (red), $p$ (blue), and (d) $w$, as functions of $R$.}
\end{figure}

To implement the JNW solution, it is more convenient to present the equation as a function of $R$ \cite{Chew:2024ggd}:
\begin{eqnarray}
\dot{R}^{2} \left( \frac{dr}{dR} \right)^{2} + V_{\mathrm{eff}}(r(R)) &=& 0,\\
\frac{d\sigma}{dR} &=& - \frac{2}{r(R)} \left( 1+w_{s} \right) \sigma \frac{dr}{dR} - \left( \sigma - \frac{\cos \chi_{0}}{4\pi r(R)} \right) \frac{d\Phi}{dR}.
\end{eqnarray}
Fig.~\ref{fig:type2} is an example of the solution. Fig.~\ref{fig:type2} (a) shows that $V_{\mathrm{eff}} < 0$ if $r \lesssim 8$, and hence, classical solution is allowed if $r \lesssim 8$. We can notice that the shell collapses and eventually forms a naked singularity.

The problem is, the energy-momentum tensor at the shell violates the null energy condition: see Fig.~\ref{fig:type2} (b), although the energy-momentum tensor of the interior region does not violate the condition (Fig.~\ref{fig:type2} (c) and (d); here, $\sigma$ is positive for $r \lesssim 8$).

\begin{figure}
\centering
\mbox{
(a)
\includegraphics[scale=0.35]{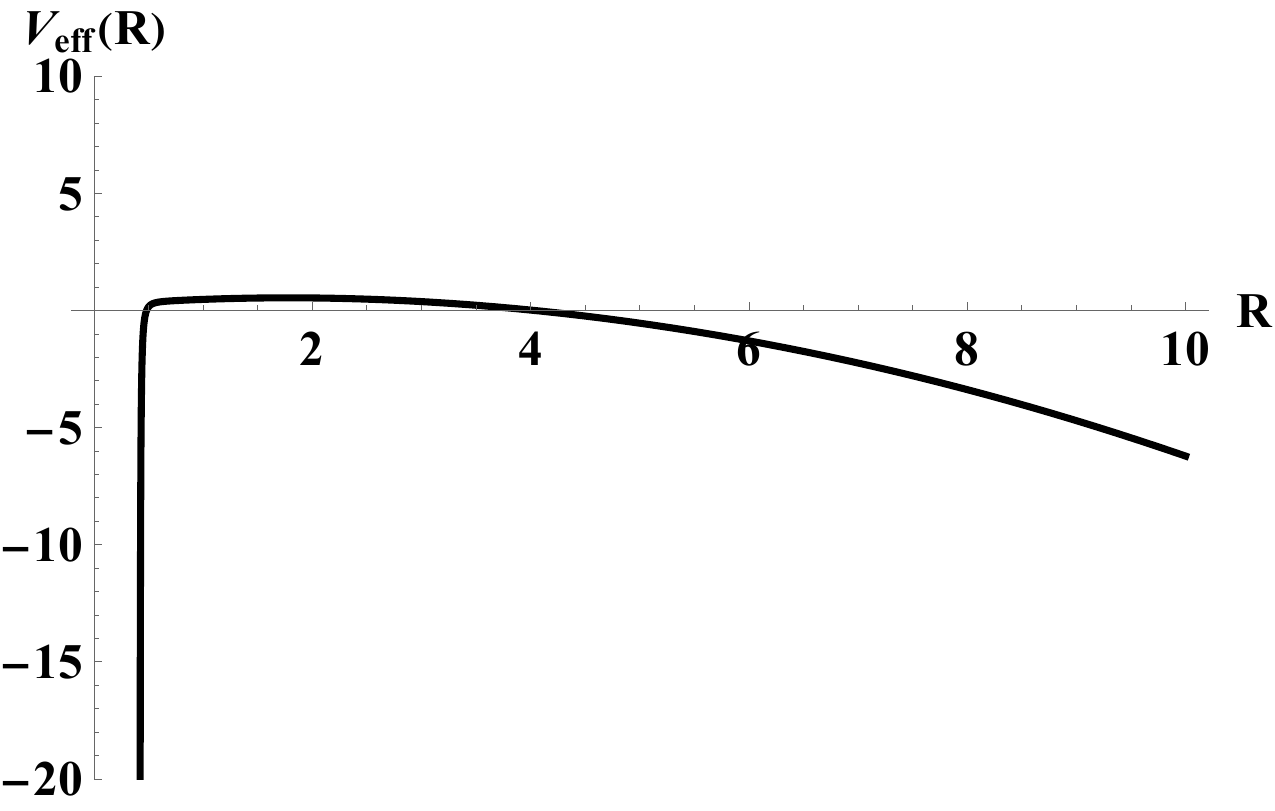}
(b)
\includegraphics[scale=0.35]{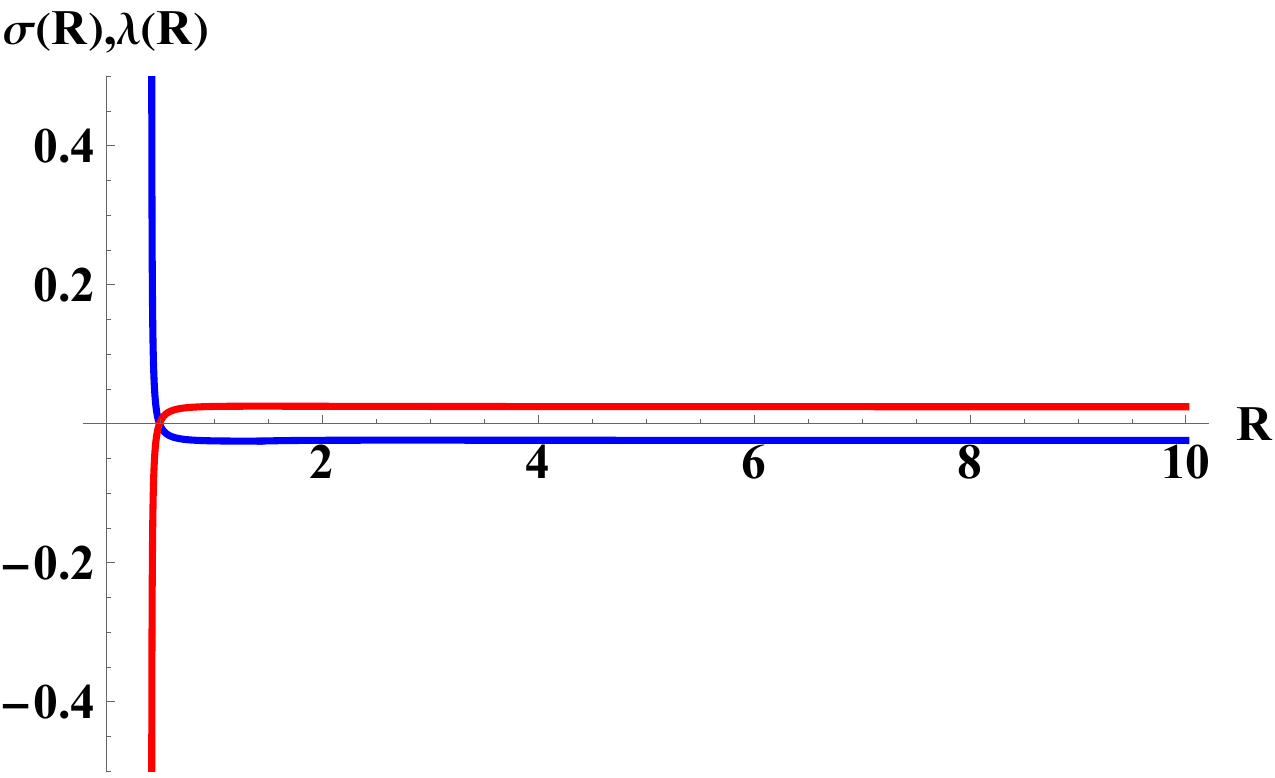}
}
\caption{\label{fig:type3}An example of the OS collapse with JNW solution ($\sigma$ diverges to negative infinity), where $A = 0.2$, $m = 1$,  $\sigma_{0} = 0.025$, $\chi_{0} = 0.1$ and $w_{s}=-1$: (a) $V_{\mathrm{eff}}$ and (b) $\sigma$ (red), $\lambda$ (blue).}
\end{figure}

When we checked the value of $\sigma$ while changing the value of parameters, we could classify two types as follows. One type is that the value of sigma diverges to negative infinity (Fig.~\ref{fig:type3}). This case violates the null energy condition as previously explained. However, one may think that a large shell satisfying the null energy condition may tunnel through the potential barrier into the singularity; after the tunneling, the null energy condition must be violated, but one can think that the shell was sufficiently pushed toward the singularity.

\begin{figure}
\centering
\mbox{
(a)
\includegraphics[scale=0.35]{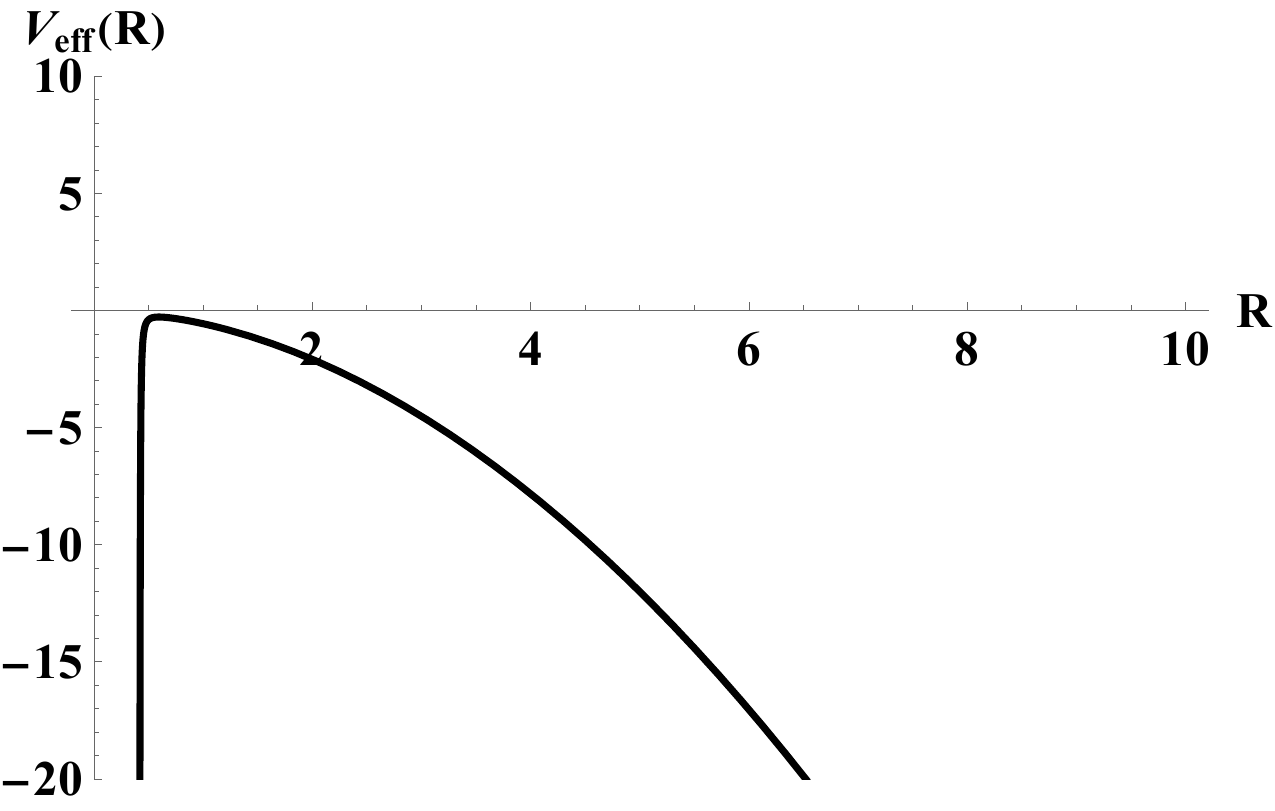}
(b)
\includegraphics[scale=0.35]{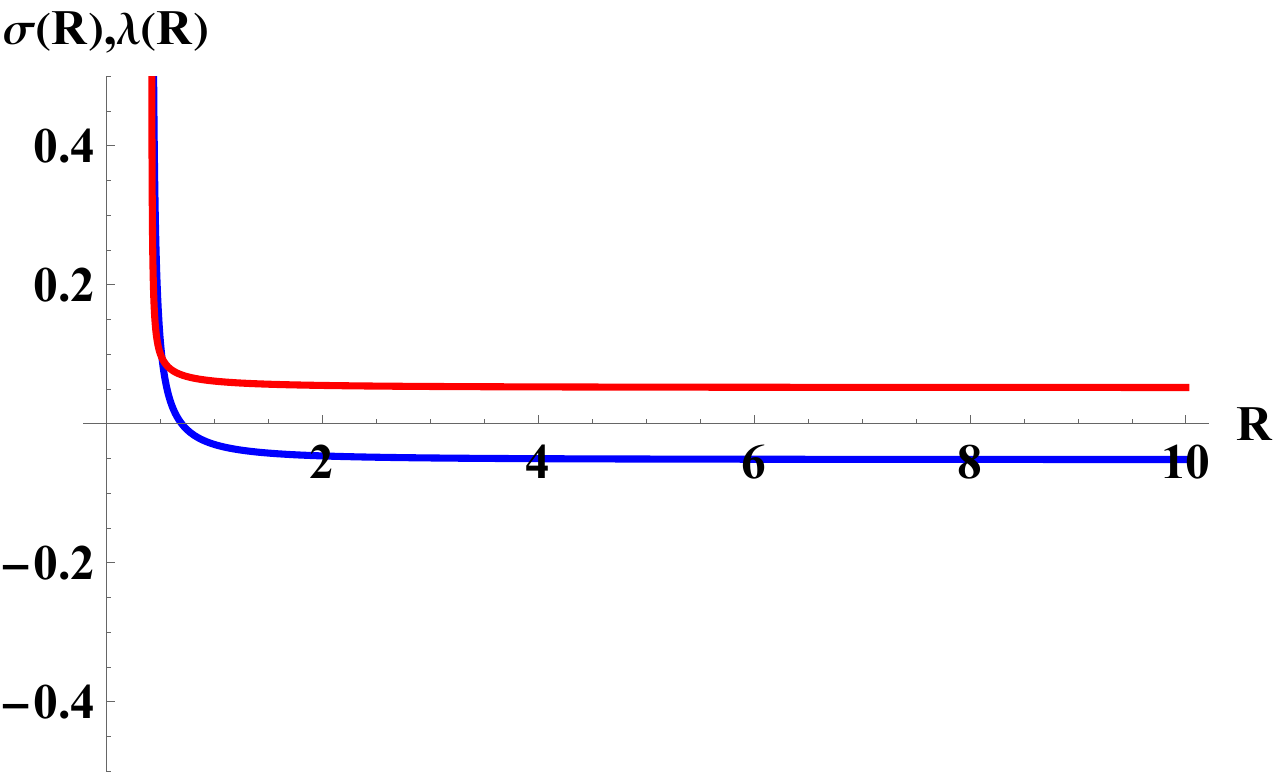}
}
\mbox{
(c)
\includegraphics[scale=0.35]{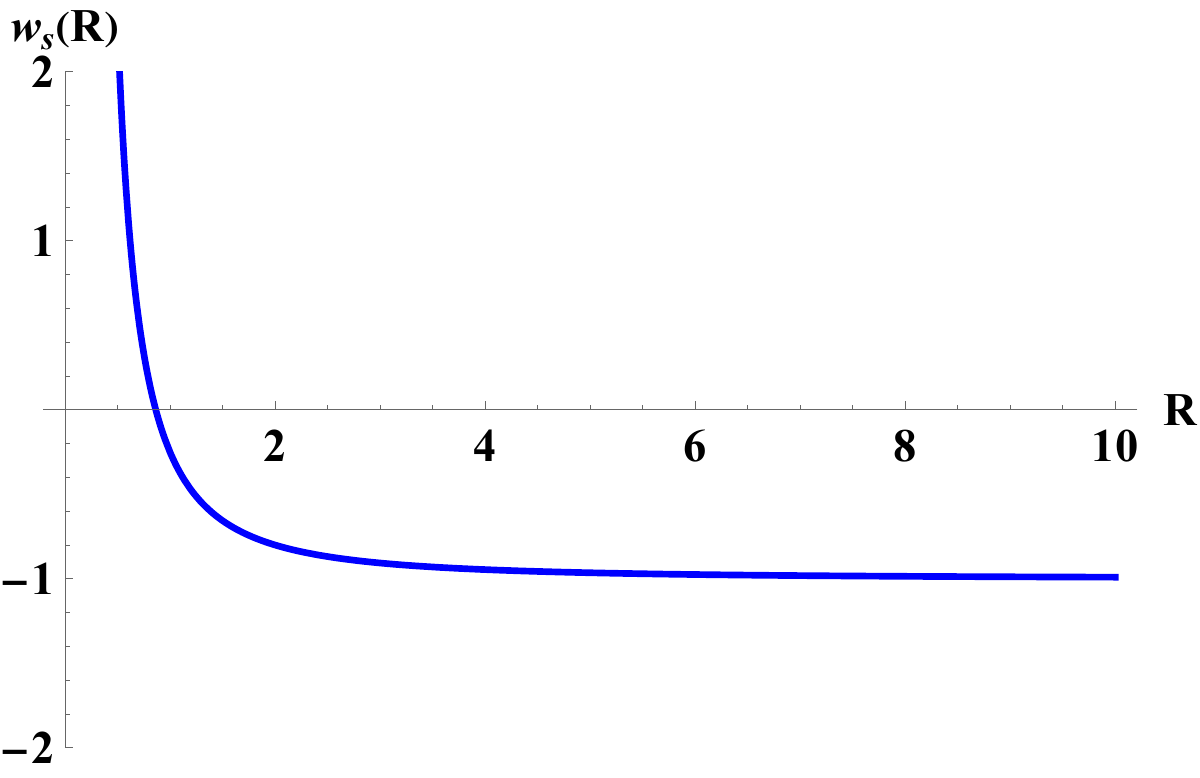}
(d)
\includegraphics[scale=0.35]{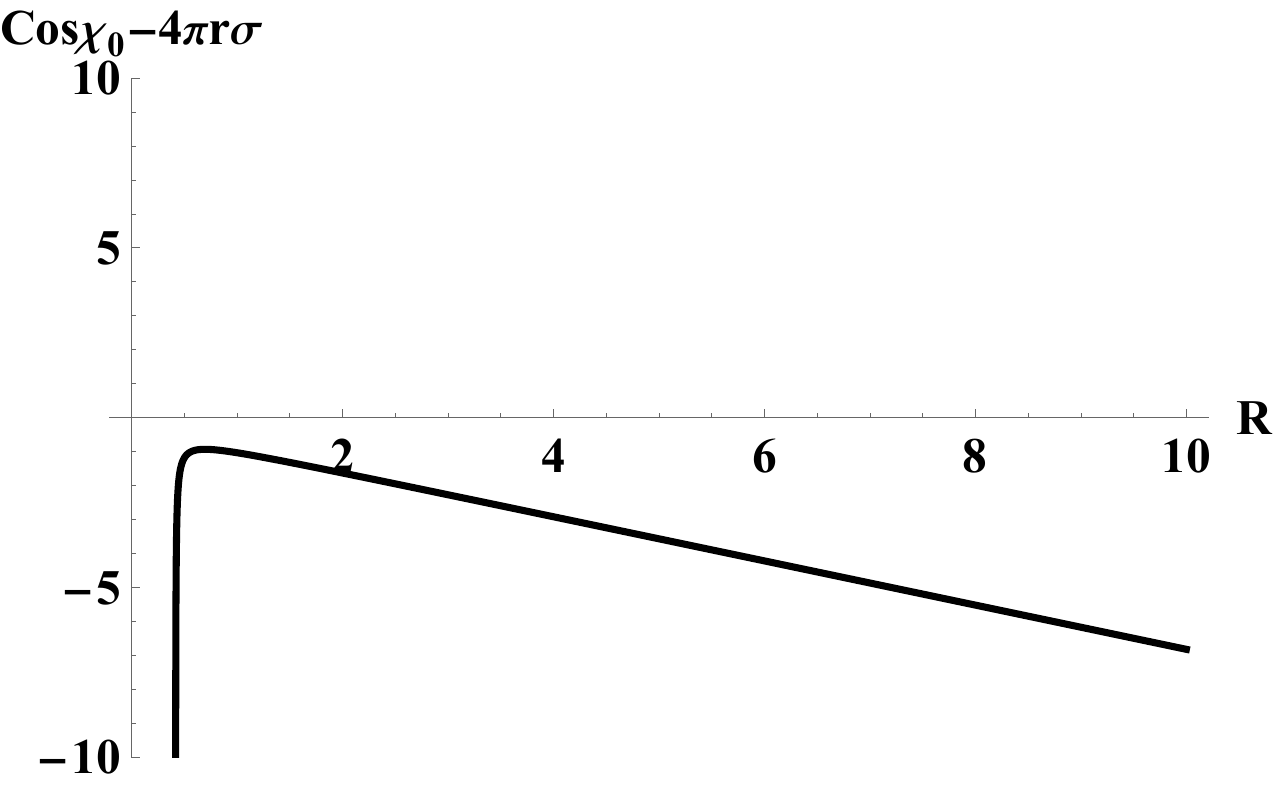}
}
\caption{\label{fig:type4}An example of the OS collapse with JNW solution ($\sigma$ diverges to positive infinity), where $A = 0.2$, $m = 1$,  $\sigma_{0} = 0.055$, $\chi_{0} = 0.1$ and $w_{s}=-1$: (a) $V_{\mathrm{eff}}$, (b) $\sigma$ (red), $\lambda$ (blue), (c) $w_{s}$, and (d) $\cos \chi_{0} - 4\pi r \sigma $.}
\end{figure}

Another case is that the value of $\sigma$ diverges to positive infinity (Fig.~\ref{fig:type4}). The formation of a naked singularity seems possible through the gravitational collapse, because the tension is always positive and the null energy condition is satisfied. However, it is not physically possible in this case because the sign of the extrinsic curvature (the sign of $\cos \chi_{0} - 4\pi r \sigma$, see Fig.~\ref{fig:type4} (d)) becomes negative.


%\begin{figure}
%\centering
%\mbox{
%(a)
%\includegraphics[scale=0.35]{9.pdf}
%(b)
%\includegraphics[scale=0.35]{12.pdf}
%}
%\mbox{
%\includegraphics[scale=0.35]{17.png}
%}
%\caption{\label{fig:type5}An example of the OS collapse with a shell assuming the JNW solution, where $A = 0.54$, $m = 1$, $\chi_{0} = 1$ : (a) $\sigma(R)$ and (b) $w_{s}(R)$.}
%\end{figure}
%
%Finally, Fig.~\ref{fig:type5} shows the result as we vary various parameters. The tension $\sigma$ must approach negative values eventually, but such a turning point that violates the null energy condition can be pushed arbitrarily close to the singularity. Hence, it is reasonable that violating the null energy condition is necessary to allow the formation of a naked singularity, but it is fair to say that the violation can be confined only near the singularity.

\subsection{Summary}

We summarize our results as follows.
\begin{itemize}
\item[--] First, we have considered several metric ansatz with a naked singularity. Such a model was used in many contexts, especially regarding regular black hole models. Using the OS collapse, it is possible to construct a naked singularity without violating the null energy condition or causality. However, the problem is that the metric ansatz itself is not a solution to the Einstein equation. 
\item[--] Second, we have considered the naked singularity as a solution of the Einstein equation and the matter field equation. Indeed, introducing an additional matter field is necessary, and the metric form naturally requires a thin-shell on top of the perfect fluid. Interestingly, the collapse of the star interior and the formation of a naked singularity is possible, but at some moment, the thin-shell must violate the null energy condition.
\end{itemize}
Therefore, we conclude that it was not possible to form a naked singularity using a gravitational collapse based on physically viable assumptions. However, there are some valuable points that we need to mention.
\begin{itemize}
\item[--] If we can construct a model that is a solution of the Einstein equation, where the metric form is $ds^{2} - f(r) dt^{2} + f^{-1}(r) dr^{2} + r^{2} d\Omega^{2}$, a naked singularity formation is possible. There might be a possibility that such a solution can be allowed, for example, in modified gravity models. Of course, if it is the case, the OS collapse model must be investigated again based on the modified gravity model.
\item[--] When considering the JNW solution as an outside solution, we reported that the thin-shell must violate the null energy condition. However, it is fair to say that we can push the moment that violates the null energy condition very close to the naked singularity. If the naked singularity is a kind of quantum gravitational object with a non-vanishing volume, we can show that the classical matter can reach sufficiently close to the object. This might be realized if we introduce quantum tunneling from a large to a small shell. This may indicate that the exact classical formation of a naked singularity is impossible, but a reasonably close approach to the singularity might be allowed.
\end{itemize}

\section{\label{sec:dis}Discussion}

This paper asks whether we can construct a physically viable gravitational collapse model that eventually forms a naked singularity. To provide the gravitational collapse process, we introduce the OS model, which describes the gravitational collapse of a perfect fluid star interior. After the collapse happens, can the resulting spacetime be a naked singularity solution?

Considering the metric ansatz that allows a naked singularity and the OS collapse without introducing a shell, it is possible to form a naked singularity without violating the null energy condition. However, we can ask whether the metric ansatz is a solution of a model or not. Therefore, in itself, this is just a toy model; however, if we can find a theoretical example that explains the metric ansatz as a solution, this can be a concrete model that violates weak cosmic censorship.

To avoid this problem, we need to find an exact solution, which requires an additional matter field. We considered a genuine solution of the Einstein equation and the matter field equation: the JNW solution. Then, using the original form of the OS collapse is impossible, and we need a thin-shell that covers the perfect fluid interior. We observed that the thin-shell should eventually violate the null energy condition. However, it is fair to say that the shell can approach the naked singularity at a close distance without violating the null energy condition. If we believe that the naked singularity is a physical object with a quantum gravitational size, we may realize an effective construction of a naked singularity from gravitational collapses. Perhaps, quantum tunneling may help and force this process, but we leave this idea for future work.

Our results provide interesting implications. First, our techniques can be used for further investigation to demonstrate physically acceptable gravitational collapse models. As a result, we revealed why it is not easy to violate cosmic censorship in classical manners. On the other hand, if we think about quantum mechanical effects, weak cosmic censorship might be violated effectively; a more physically correct explanation is that the shell approaches the singularity and goes beyond the scope where general relativity is valid. Perhaps this might be the best effort that we can do to violate weak cosmic censorship.

Can we do further in a classical way? Can a modified gravity model provide an example that violates weak cosmic censorship? What will a naked singularity's effect be in the real world, if it exists? We leave these interesting topics for future investigation.

\section*{Acknowledgment}
This work was supported by the Financial Supporting Project of Long-term Overseas Dispatch of PNU's Tenure-track Faculty, 2025.

\end{document}